\documentclass[aps,prd,preprint]{revtex4-1}
\usepackage{epsfig}
\usepackage{graphicx}

\begin{document}
\title{Influence of a uniform magnetic field on dynamical chiral symmetry breaking in QED$_3$}
\author{Jian-Feng Li$^{1,2}$, Hong-Tao Feng$^{3}$, Yu Jiang $^{4}$, Wei-Min Sun$^{2,5}$ and Hong-Shi Zong$^{2,5}$}
\address{$^{1}$ College of Mathematics and Physics, Nantong University, Nantong 226007, China}
\address{$^{2}$ Department of Physics, Nanjing University, Nanjing 210093, China}
\address{$^{3}$ Department of Physics, Southeast University, Nanjing 211189, China}
\address{$^{4}$ Center for Statistical and Theoretical Condensed Matter Physics, Zhejiang Normal University, Jinhua 321004, China}
\address{$^{5}$ Joint Center for Particle, Nuclear Physics and Cosmology, Nanjing 210093, China}
\begin{abstract}

We study dynamical chiral symmetry breaking (DCSB) in an effective
QED$_{3}$ theory of $d$-wave high temperature cuprate
superconductors under a uniform magnetic field. At zero temperature,
the external magnetic field induces a mixed state by generating
vortices in the condensate of charged holons. The growing magnetic
field suppresses the superfluid density and thus reduces the gauge
field mass which is opened via the Anderson-Higgs mechanism. By
numerically solving the Dyson-Schwinger gap equation, we show that
the massless fermions acquires a dynamical gap through DCSB
mechanism when the magnetic field strength $H$ is above a critical
value $H_{c}$ and the fermion flavors $N$ is below a critical value
$N_{c}$. Further, it is found that both $N_{c}$ and the dynamical
fermion gap increase as the magnetic field $H$ grows. It is expected
that our result can be tested in phenomena in high temperature
cuprate superconductors.

\bigskip

Key-words: QED$_3$; dynamical chiral symmetry breaking; gap
equation; $d$-wave superconductor
\bigskip

E-mail: zonghs@chenwang.nju.edu.cn. ~~~~PACS Numbers: 11.10.Kk, 11.15.Tk, 11.30.Qc

\end{abstract}
 \maketitle

Analogous to Quantum Chromodynamics (QCD), Quantum Electrodynamics
in (2+1) spacetime dimensions (QED$_3$) has some interesting
features, such as dynamical chiral symmetry breaking (DCSB) in the
massless fermion limit and confinement
\cite{a1,a2,a3,a4,a5,a6,a7,a8,a9}. Therefore QED$_3$ can be
regarded as a useful model by which one can gain important insights
into the aspects of QCD. In addition, QED$_3$ and its variants also
have proved to be the effective field theories for some
two-dimensional condensed matter systems, including high temperature
superconductors \cite{a10,a11,a12,a13,a14,a15,a16,a17,a18} and
graphene \cite{Khveshchenko,Khveshchenko1,Gusynin, a19,a20,a21, Liu09, Gamayun,Kotov}. Because of these features, QED$_3$ has been extensively
studied in recent years.

DCSB occurs when the massless fermions appearing in a Lagrangian
that respects chiral symmetry acquire a nonzero mass through
nonperturbative effects at low energy. In a four-fermion interaction
model, Nambu and Jona-Lasinio \cite{a22} first transplanted the
fermion pairing idea in the BCS theory of superconductivity into
particle physics to generate a fermion mass without assuming a
Yukawa type coupling between fermion and Higgs scalar field (using
Yukawa type coupling). In Refs. \cite{a3,a4} Appelquist \emph{et}
\emph{al.} first studied DCSB in massless QED$_3$ with $N$ fermion
flavors at zero temperature by solving the Dyson-Schwinger equations
(DSEs) for fermion self-energy in the lowest-order of $1/N$
expansion, and found that a fermion mass was dynamically generated
when $N$ is less than a critical number $N_c = 32/\pi^2$. Later,
Nash further considered higher order corrections, and showed that
the critical number of fermion flavor still exists and $N_{c} =
\frac{128}{3\pi^{2}}$ \cite{a5}. At finite temperature, Aitchison et
al. used an approximate treatment of DSEs for the fermion
self-energy in the $1/N$ expansion \cite{a23} and found that $N_c$
is temperature-dependent and chiral symmetry is restored above a
certain critical temperature. Further analogy with QCD at finite
temperature and chemical potential has been explored
\cite{He,Feng1,Feng2,Feng3}. The methods described there could be
employed to extend the study of QED$_3$ at nonzero temperature and
chemical potential and explore whether and under which condition
chiral symmetry restoration and deconfinement are coincident.
Recently, the authors in Ref. \cite{bashir} studied the impact of a
uniform magnetic field on explicit chiral symmetry breaking in
QED$_3$.

Besides its interests in high energy physics, QED$_3$ also turned
out to be an effective low-energy theory of high temperature cuprate
superconductors. In the past twenty years, it has been extensively
applied to understand some unusual physics of high temperature
superconductors \cite{a13}. In two early papers \cite{Affleck,Ioffe}, it
was shown that the half-filled state on the CuO$_2$ plane of undoped
cuprates can be mapped to a (2+1)-dimensional continuum field theory
composed of massless Dirac fermions, called spinons, and U(1) gauge
field, which is exactly QED$_3$. In this effective theory, the
physical fermion flavor is $N=2 < N_c$, so DCSB can take place at
zero temperature. Physically, DCSB corresponds to the formation of
long-range antiferromagnetic (AFM) order \cite{Kim97,Kim99}. At
finite doping concentration, the simple QED$_3$ of Dirac fermions is
not sufficient to model the dynamics of doped holes. To overcome
this difficulty, Lee and Wen developed an SU(2) gauge field theory
to describe the underdoped cuprates \cite{a13}. In the so-called
staggered flux phase, two components of the SU(2) gauge field are
gapped and thus can be neglected, leaving only one massless
component of gauge field \cite{a13}. However, in underdoped
cuprates, there appear additional scalar boson excitations, called
holons, which describes the dynamics of doped holes. Therefore, the
effective continuum theory of underdoped cuprates is an extended
QED$_3$ theory of massless spinons and scalar holons \cite{Kim97,
Kim99}. The spinons and holons both interact with the gauge field
strongly, but there is no direct coupling between them. This theory
can exhibit a lot of interesting features, and has been investigated
extensively in the past several years \cite{a13, Affleck,Ioffe, Kim97,
Kim99,  Rantner2}.

Within this effective theory, superconductivity is achieved by the
condensation of holons at low temperature. In the superconducting
(SC) state, the gauge field acquires a finite mass gap through the
famous Anderson-Higgs (AH) mechanism. As shown in Ref. \cite{Liu}, this
gauge field mass weakens the gauge interaction and thus rapidly
suppresses DCSB. If we use DCSB to represent AFM order and use gauge
field mass to represent SC order, then the competition between AFM
order and SC order can be understood by the competition between DCSB
and AH mechanism \cite{Liu}.

Now suppose we introduce an external magnetic field into the SC
state, then a mixed state, or called vortex state, emerges since
high temperature superconductors are known to be extreme type-II
superconductors. The most important influence of external magnetic
field is to suppress the superfluid density of the SC state. From a
field-theoretic point of view, this implies the suppression of the
gauge field mass, which is proportional to the superfluid density.
Since the gauge field mass can have important influence on DCSB, it
is interesting to study the fate of DCSB when the strength of
external magnetic field is varied. The main purpose of this paper is
to study the impact of the external magnetic field on the critical
fermion flavor $N_{c}$.

In this paper, we first employ the DSEs method to construct the
fermion gap equation at zero temperature after including the effect
of an external uniform magnetic field. Then we solve the gap
equation to study the influence of external magnetic field on DCSB
in the mixed state of a d-wave cuprate superconductor. The main
result of our paper is that both the critical fermion flavor $N_{c}$
and the gap increase with growing magnetic field.

It is well known that QED$_3$ is super-renormalizable and has an
intrinsic mass scale given by the coupling constant
$\alpha=\frac{Ne^{2}}{8}$. In Euclidean space, the Lagrangian of massless QED$_3$  with $N$ fermion flavors reads
\begin{equation}\label{eq1}
\mathcal{L}=\sum^N_{i=1}\bar\psi_i(\not\!\partial+\mathrm{i}e\not\!a)\psi_i+\frac{1}{4}F^2_{\rho\nu}+\frac{1}{2\xi}(\partial_\rho\
a_\rho)^2,
\end{equation}
where the $4 \times1$ spinor $\psi_i$ represents the fermion field,
$i=1,\cdots,N$ are the flavor indices, and $\xi$ is the gauge
parameter. Here, the massless Dirac fermions are not the
electron-like excitations, but are actually the gapless spinons
excited from the nodes of $d$-wave energy gap. These fermions have
spin $1/2$ but are electrically neutral. The gauge field $a_{\mu}$ is also not the usual vector potential of electromagnetic field. Instead, it
emerges as the consequence of strong correlation between electrons
on the CuO$_2$ planes of high-$T_c$ superconductors. This effective
field theory can be derived from the microscopic $t$-$J$ model of
high-$T_c$ superconductors with the help of slave-boson technique
\cite{a13}.

The DSEs for the fermion propagator $G(p)$ at zero temperature reads
\begin{equation}\label{eq2}
G^{-1}(p)=i\gamma\cdot p +e^2\int \frac{d^3k}{(2\pi)^3}\gamma_\rho
G(k)\Gamma_\nu(p,k)D_{\rho\nu}(q),
\end{equation}
where $D_{\rho\nu}$ is the propagator of the $a_{\mu}$ field and $q=p-k$. For simplicity, in this paper we adopt the rainbow approximation,
i.e., we approximate the full vertex function $\Gamma_\nu$(p,k) with
the bare vertex $\gamma_{\nu}$ and we neglect the wave function
renormalisation (this is a commonly employed approximation in the study of fermion propagator of QED$_3$ at finite temperature \cite{Khveshchenko,a24,a25}). Similar with the usual DSE study, in the following we will work in Landau gauge.

At finite temperature $T$ the DSEs
for the fermion propagator is given by
\begin{equation}\label{eq3}
G^{-1}(p)=G_{0}(p)-\frac{e^{2}}{\beta}\sum^{\infty}_{n=-\infty}\int\frac{d^{2}k}{(2\pi)^{2}}\gamma^{\mu}G(k_{0},K,\beta)\Delta_{\mu\nu}(q_{0},Q,\beta)\gamma^{\nu},
\end{equation}
where
\begin{equation}\label{eq4}
G(p)=\frac{1}{ip_{0}\gamma_{0}-v_{F}p_{1}\gamma_{1}-v_{\Delta}p_{2}\gamma_{2}
- m(p^2)}
\end{equation}
is the fermion propagator at finite temperature and $m(p)$ represents the Dirac fermion mass. Here $v_{F}$ is the Fermi velocity, $v_\Delta$ is the gap velocity. Since the velocity anisotropy is unimportant to the critical
behavior \cite{a18}, we can set $v_{F}=v_{\Delta}=1$, and
\begin{eqnarray}\label{eq5}
p=(p_{0},\textbf{p}),P=|\textbf{p}|,p_{0}=(2m+1)\frac{\pi}{\beta},\nonumber\\
k=(k_{0},\textbf{k}),K=|\textbf{k}|,k_{0}=(2n+1)\frac{\pi}{\beta},\nonumber\\
q=(q_{0},\textbf{q}),Q=|\textbf{q}|=|\textbf{p}-\textbf{k}|,q_{0}=2(m-n)\frac{\pi}{\beta}.
\end{eqnarray}
Under the above approximation the inverse of the fermion propagator can be written as
\begin{equation}\label{eq6}
G^{-1}(p)=i\gamma\cdot p +m(p^{2}).
\end{equation}
Taking the trace of both sides of Eq. (3) yields a closed integral equation for
m(p) \cite{a23,a24,a25}:
\begin{equation}\label{eq7}
m(p^2)=\frac{\alpha}{N\beta}\sum^{\infty}_{n=-\infty}\int\frac{d^{2}k}{(2\pi)^{2}}\Delta(q_{0},Q,\beta)\frac{m(k^2)}{k^{2}+m^{2}(k^2)}
\end{equation}
with
\begin{equation}\label{eq8}
\Delta(q_{0},Q,\beta)=\frac{1}{8}Tr[\gamma_{\mu}\Delta_{\mu\nu}(q_0,Q,\beta)\gamma_{\nu}],
\end{equation}
where
\begin{equation}\label{eq9}
\Delta_{\mu\nu}(q_0,Q,\beta)=\frac{\delta_{\mu3}\delta_{\nu3}}{Q^{2}+\Pi_{0}(Q,\beta)+m_{a}^{2}}
\end{equation}
and
\begin{equation}\label{eq10}
\Pi_{0}(Q,\beta)=\frac{\alpha}{8\beta}\left[Q\beta+\frac{16\ln2}{\pi}\exp(-\frac{\pi\beta}{16\ln2})\right].
\end{equation}
When the photon acquires a mass $m_a$ via the Anderson-Higgs mechanism,
the massless photon propagator is replaced by the massive photon
propagator (9). So we have
\begin{eqnarray}\label{eq11}
m(p^2)=\frac{\alpha}{8N\beta}\sum_{n=-\infty}^{\infty}\int\frac{d^{2}k}{(2\pi)^{2}}\frac{m(k^2)}{[K^{2}+m^{2}(k^2)+\varpi_{n}^{2}][Q^{2}+\Pi_{0}(\beta,Q)+m_{a}^{2}]}.
\end{eqnarray}
The summation over infinite Matsubara frequencies can be done
analytically:
\begin{eqnarray}\label{eq12}
&&\sum_{n=-\infty}^{\infty}\frac{1}{\varpi_{n}^{2}+K^{2}+m^{2}(k^2)}\nonumber\\
&&=\frac{1}{2\sqrt{K^{2}+m^{2}(k^2)}}
\sum_{n=-\infty}^{\infty}\left[\frac{1}{i\varpi_{n}+\sqrt{K^{2}+m^{2}(k^2)}}-\frac{1}{i\varpi_{n}-\sqrt{K^{2}+m^{2}(k^2)}}\right]\nonumber\\
&&=\frac{\beta}{2\sqrt{K^{2}+m^{2}(k^2)}}\tanh\left[\frac{\beta}{2}\sqrt{K^{2}+m^{2}(k^2)}\right].
\end{eqnarray}
Finally we obtain the temperature- and  momentum-dependent gap
function
\begin{equation}\label{eq13}
m(p^2)=\frac{\alpha}{8 N\pi^{2}}\int
d^{2}K\frac{m(k^2)}{\sqrt{K^{2}+m^{2}(k^2)}}\frac{\tanh[\frac{\beta}{2}\sqrt{K^{2}+m^{2}(k^2})]}{Q^{2}+\Pi_{0}(\beta,Q)+m_{a}^{2}}.
\end{equation}
Here we note that in the limit of zero temperature $\beta\rightarrow\infty$, we have
\begin{equation}\label{eq14}
\Pi_{0}(Q,\beta)\rightarrow\frac{\alpha Q}{8},
\end{equation}
and the gap equation further simplifies into
\begin{equation}\label{eq15}
m(P^2)=\frac{\alpha}{8 N\pi^{2}}\int
d^{2}K\frac{m(K^2)}{\sqrt{K^{2}+m^{2}(K^2)}}\frac{1}{Q^{2}+\frac{\alpha
Q}{8}+m_{a}^{2}}.
\end{equation}

In order to see how the gauge field acquire a mass $m_{a}$, we now
introduce the additional interaction term between gauge field and scalar boson field
\begin{eqnarray}
\mathcal{L}_{h} = \phi^{\ast}(\partial_0 + ia_0)\phi -
\frac{1}{2m_h}\phi^{\ast}(\mathbf{\partial} + i\mathbf{a})^2\phi +
\frac{u}{2}\phi^2 + \frac{\beta}{4}\phi^4,
\end{eqnarray}
which is so-called Abelian Higgs model. The scalar field $\phi$
represents the bosonic holons, which has spin-$0$ and carry charge
$-e$ \cite{a13}. This Lagrangian describes the motion of the charge
degrees of freedom of electrons on the CuO$_2$ planes of underdoped
cuprate superconductors \cite{a13}. Unfortunately, there is
currently not a clear understanding of the dynamics of holons . In
previous treatments of this model, the holon sector was simply
neglected \cite{Kim97, Kim99, Rantner2} in the analysis of the
interaction between Dirac fermions and gauge field. In this paper,
we believe that the most prominent effect of this additional
interaction is to generated a finite mass for the gauge field in the
SC state. When $u> 0$, the system stays in the normal state and the vacuum
expectation value of boson field $\langle \phi \rangle = 0$, so the
Lagrangian respects the local gauge symmetry. When $u < 0$, the
system enters the SC state and the boson field develops a finite
expectation value $\langle \phi \rangle \neq 0$, then the local
gauge symmetry is spontaneously broken. According to the
Anderson-Higgs mechanism, the gauge field acquires a finite mass
$m_a$ after absorbing the massless Goldstone boson generated due to
spontaneous breaking of continuous gauge symmetry. The finite gauge
field mass is able to characterize the achievement of
superconductivity. Actually, this mass is proportional to the
superfluid density of the SC state, $m_a \propto \langle \phi
\rangle$.

The external magnetic field turns the uniform SC state to a mixed
state, where the superfluid density and therefore the gauge field
mass decreases with growing field strength $H$. If we can get a
relationship between gauge field mass $m_a$ and field strength $H$,
we will be able to study the impact of magnetic field on DCSB. Since
the holon dynamics is poorly understood, we will use
phenomenological arguments.

As mentioned above, DCSB can describe the AFM order and the gauge
boson mass can describe the SC order \cite{a11, Liu}. Based on these
correspondences, we are able to give a qualitatively physical
description of the competition between AFM order and SC order in the
mixed state of high-$T_{c}$ superconductors. In the absence of
external magnetic field, the system stays in a uniform SC state at
zero temperature, with a superfluid density $\Lambda_{s}$. As an
external magnetic field is introduced, the phase coherence among
holons is disputed and there appear single vortices inside which
superconductivity is destroyed. In this mixed state, the superfluid
density becomes inhomogeneous and takes different value at different
positions. However, after averaging over the vortices, it is still
possible to obtain a homogeneous superfluid density
$\Lambda_{s}(H)$, which is a function of magnetic field. On physical
grounds, it is easy to suppose that $\Lambda_{s}(H)$ is a decreasing
function of field strength $H$. The gauge boson mass in the mixed
state is proportional to the superfluid density, so $m_a =
a\Lambda_{s}(H)$, where $a$ is a dimensionless constant which
depends on the doping concentration. Now we have a clear picture: as
magnetic field $H$ grows, the superfluid density $\Lambda_{s}(H)$
decreases, and hence the gauge boson mass $m_a$ decreases, which
will raise the critical fermion flavor $N_c$ and may finally induce
DCSB in the mixed state. Once a finite fermion mass is dynamically
generated, its magnitude $m(p^2)$ will be an increasing function of
magnetic field $H$. In the following we will verify this qualitative
argument by numerically solving the gap equations in the presence of
an external magnetic field.

To study the gap equation, the superfluid density $\Lambda_{s}(H)$
can be obtained by averaging over the vortices. In the study of
mixed state, Volovik \cite{a27} at first proposed a semiclassical
approach and showed that the density of states goes as $\sqrt{H}$ at
low temperatures, which has been observed by experiments \cite{a28}.
A more complicated computation of superfluid density
$\Lambda_{s}(H)$ within the semi-classical approximation has already
been performed by Sharapov \emph{et} \emph{al.} \cite{a29}. The
superfluid stiffness is given by
\begin{equation}\label{eq17}
\Lambda_{s}^{ij}(T,H)=\tau^{ij}-\Lambda_{n}^{ij}(T,H),
\end{equation}
where $\tau_{ij}$ is the diamagnetic tensor and $\Lambda_{n}$
represents the normal fluid density divided by the carrier mass.
After taking some approximations and performing some complicated
calculations, Sharapov \emph{et} \emph{al.} obtained \cite{a29}
\begin{equation}\label{eq18}
\Lambda_n(H) = \frac{v_F}{\pi
v_{\Delta}}\frac{E_{H}^{2}}{\sqrt{E_{H}^{2}+\Delta^{2}_{d_{xy}}}}.
\end{equation}
When the superconducting gap $\Delta _{d_{xy}}(H)\ll\sqrt{H}$, we can
neglect the $\Delta _{d_{xy}}$ and $\Lambda_n(H)$ still has the
behavior $\sim \sqrt {H}$. The superfluid density then becomes
\begin{equation}\label{eq19}
\Lambda_s(H) = \tau - \frac{v_F}{\pi v_{\Delta}}E_{H}.
\end{equation}
Here, $\tau$ is the superfluid density at zero temperature in the
absence of magnetic field and the ratio between $v_{F}$ and
$v_{\Delta}$ appears as a coefficient. When $v_{F}$ and $v_{\Delta}$
multiply particle momenta, the anisotropy is irrelevant, but the
anisotropy is very important in this expression. In the following
discussions, as in Ref. \cite{a29}, we simply set the ratio
$v_{F}/v_{\Delta}= 30$, and at zero temperature the superfluid
density is taken to be $\tau=1500 \mathrm{K}$ and the energy scale
$E_{H}[K] = 38 \mathrm{K}\mathrm{T}^{-1/2}\sqrt{H[T]}$. Since the
gauge boson $m_{a}$ is proportional to superfluid density, we use
the ansatz that $m_{a} = a\Lambda_{s}(H)$, where parameter $a$ is a
free variable.

After obtaining the effective gauge boson propagator (15), we can
now obtain the fermion gap equation in the presence of an external
magnetic field
\begin{equation}\label{eq20}
m(P^2)=\frac{\alpha}{8 N\pi^{2}}\int
d^{2}K\frac{m(K^2)}{\sqrt{K^{2}+m^{2}(K^2)}}\frac{1}{Q^{2}+\frac{\alpha
Q}{8}+[a\Lambda_{s}(H)]^{2}}.
\end{equation}
The iteration algorithm can be employed to numerically solve this
equation. In our numerical computations, we set $\alpha = 1$ so that
each quantity with mass dimension is re-scaled and becomes
dimensionless. In the real numerical calculations, we adopt an
ultraviolet cutoff $\Lambda=10^{5}$, which is large enough to ensure
that the calculated results are stable with respect to $\Lambda$. We
consider $a = 0.07, 0.1, 0.15$. For each parameter $a$, the relation
between the critical fermion flavor $N_{c}$ and the magnetic field
$H$ is shown in Fig. 1. The critical number of fermion flavor
$N_{c}$ separates the chiral symmetric phase ($m=0$ for $N>N_{c}$)
from the DCSB phase ($m>0$ for $N<N_{c}$). From Fig.1 it can be seen
that for fixed value of the parameter $a$, $N_{c}$ increases as the
magnetic field $H$ grows. For a fixed magnetic field $H$, the
smaller is the parameter $a$, the easier is it to generate an energy
gap by the mechanism of DCSB. For the physical flavor $N=2$, when
the magnetic field $H$ is large enough, a finite mass gap is
generated in the mixed state. It is obvious that the dynamical fermion mass will affect all observable physical quantities. It is expected that our result can be tested in phenomena in high temperature cuprate superconductors.

We now take the physical fermion flavor $N = 2$, and numerically
solve the gap equation by iteration method for the specific case
$a=0.07$. The momentum dependence of the fermion gap is plotted in
Fig. 2. From Fig. 2, it can be seen that the dynamically generated
mass $m(p^{2})$ is almost constant for small momenta, but decreases
down to nearly zero at large momenta for a fixed magnetic field.
This is easy to understand because the QED$_3$ theory is
asymptotically free. Another important feature is that fermion gap
$m(H)$ is an increasing function of the external magnetic field $H$.
Apparently, the DCSB that is suppressed by finite gauge boson mass
is restored when the magnetic field $H$ becomes sufficiently large.
When the fermion gap becomes larger with magnetic field, the thermal
conductivity $\kappa \propto \frac{\Gamma_0^2}{m^2(H) +
\Gamma_0^2}$, where $\Gamma_0$ is the disorder scattering rate,
becomes smaller. There will be a thermal metal-to-insulator
transition with growing magnetic field in the limit of $m(H) \gg
\Gamma_0$. These results qualitatively agree with the conclusion
given by Jiang \emph{et} \emph{al.} in \cite{a26}, where the authors
introduced a two-dimensional Coulomb interaction between Dirac
fermions to generate an excitonic gap.

\begin{figure}[htp!]
\includegraphics[width=10cm]{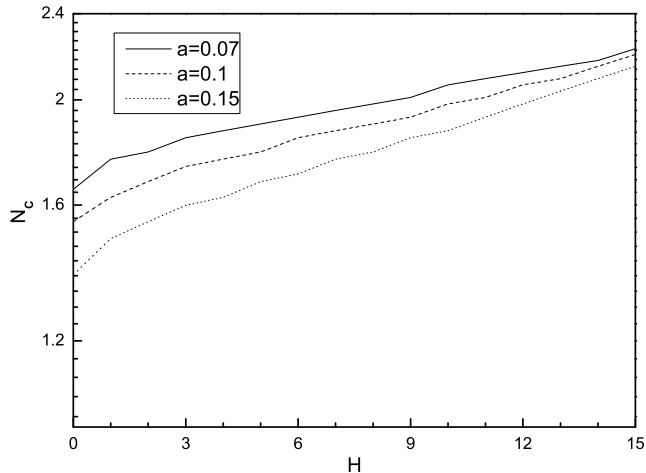}
\caption{The dependence of the critical number of fermion flavors
$N_{c}$ on the magnetic field for several values of $a$}\label{FIG1}
\end{figure}

\begin{figure}[htp!]
\includegraphics[width=10cm]{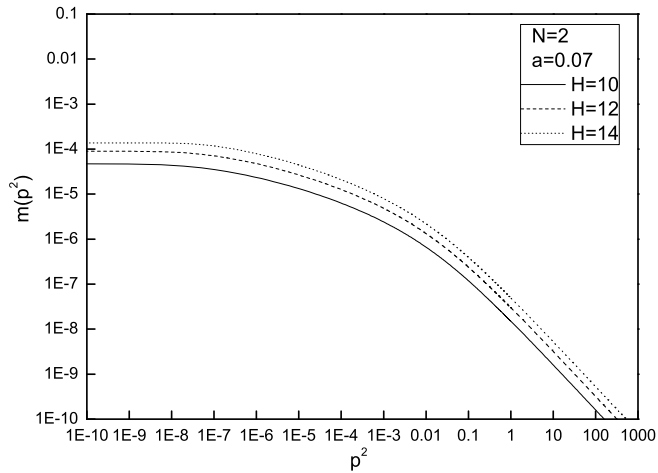}
\caption{Scaled mass gap versus scaled momentum at various magnetic
field for $N=2 $ and $a=0.07$}\label{FIG2}
\end{figure}

We note that QED$_3$ in a uniform magnetic field was studied
previously in several papers \cite{Gusynin94,Gusynin95, Ferrer1,Ferrer2}. A magnetic catalyst phenomenon was found there, which states that
DCSB can be significantly catalyzed by an external magnetic field.
In the present paper, this effect is not considered. This is mainly for two
reasons. First, the gauge field in Eqs. (1) and (16) denoted by $a_{\mu}$, is not the electromagnetic gauge field, but some new internal gauge field. In this sense the fermion does carry a gauge charge which is not the conventional electronic charge. Moreover, unlike the ordinary electrons that carry both spin and electric charge, the Dirac fermions appearing in our model are
neutral in electric charge, so they are not directly coupled to the
external magnetic field via the minimal coupling principle. Second,
a crucial assumption of the magnetic catalyst phenomenon is that the
external magnetic field induces discrete Landau levels of the Dirac
fermions \cite{Gusynin94, Ferrer1,Ferrer2}. However, it was found that there
are no Landau levels in the mixed state of high temperature superconductor \cite{Melnikov,Franz99}.

To summarize, there is a long history to solve the Dyson-Schwinger
gap equation in QED$_3$. It is interesting to apply it to some
two-dimensional condensed matter systems. In this paper we apply the
analysis to chiral symmetry breaking in high temperature cuprate
superconductors under a uniform magnetic field. It is shown that
chiral symmetry breaking takes place when the magnetic field is
above a critical value $H_c$ and the fermion flavors is below a
critical value $N_c$. Though the well-known magnetic catalysis does
not occur in this model, the magnetic field leads to chiral symmetry
breaking indirectly by suppressing the gauge boson mass. It is
expected that our result can be tested in phenomena in high
temperature cuprate superconductors.

\bigskip

\textbf{Acknowledgement}

\bigskip
We thank G.Z.Liu for helpful correspondence.
This work was supported in part by the National Natural
Science Foundation of China (under Grant Nos 10175033, 10475057,
10575050), the Research Fund for the Doctoral Program of Higher Education 
(under Grant No 200802840009) and a project funded by the Priority Academic Program Development of Jiangsu Higher Education Institution.

\end{document}